\documentclass[aps,superscriptaddress,showpacs,floatfix,amsmath,amssymb,twocolumn,preprintnumbers,]{revtex4}

\usepackage{epsfig}
\usepackage{dcolumn}
\usepackage{bm}
\usepackage{multirow}

\begin{document}

\title{Excited states in the neutron-rich nucleus $^{25}$F}

\author{Zs. Vajta}
\affiliation{Institute for Nuclear Research,
Hungarian Academy of Sciences, P.O. Box 51, Debrecen, H-4001, Hungary}
\author{M.\ Stanoiu}
\affiliation{IFIN-HH, P. O. Box MG-6, 76900 Bucharest-Magurele, Romania}
\author{D.\ Sohler}
\affiliation{Institute for Nuclear Research,
Hungarian Academy of Sciences, P.O. Box 51, Debrecen, H-4001, Hungary}
\author{G.~R.~Jansen}
\affiliation{Department of Physics and Astronomy, University of
Tennessee, Knoxville, TN 37996, USA}
\affiliation{Physics Division, Oak Ridge National Laboratory,
Oak Ridge, TN 37831, USA}
\author{F.\ Azaiez}
\affiliation{Institut de Physique Nucl\'eaire, IN2P3-CNRS,
F-91406 Orsay Cedex, France}
\author{Zs. Dombr\'adi}
\affiliation{Institute for Nuclear Research,
Hungarian Academy of Sciences, P.O. Box 51, Debrecen, H-4001, Hungary}
\author{O.~Sorlin}
\affiliation{Grand Acc\'el\'erateur National d'Ions Lourds, CEA/DSM-CNRS/IN2P3, 
B.P. 55027, F-14076 Caen Cedex 5, France}
\author{B.~A.~Brown}
\affiliation{National Superconducting Cyclotron Laboratory and Department of Physics and Astronomy, Michigan
  State University, East Lansing, MI 48824-1321, USA}
\author{M.\ Belleguic}
\affiliation{Institut de Physique Nucl\'eaire, IN2P3-CNRS,
F-91406 Orsay Cedex, France}
\author{C.\ Borcea}
\affiliation{IFIN-HH, P. O. Box MG-6, 76900 Bucharest-Magurele, Romania}
\author{C.\ Bourgeois}
\affiliation{Institut de Physique Nucl\'eaire, IN2P3-CNRS,
F-91406 Orsay Cedex, France}
\author{Z.~Dlouhy}
\affiliation{Nuclear Physics Institute, AS CR, CZ 25068, Rez, Czech Republic}
\author{Z.~Elekes}
\affiliation{Institute for Nuclear Research,
Hungarian Academy of Sciences, P.O. Box 51, Debrecen, H-4001, Hungary}
\author{Zs.~F\"ul\"op}
\affiliation{Institute for Nuclear Research,
Hungarian Academy of Sciences, P.O. Box 51, Debrecen, H-4001, Hungary}
\author{ S.~Gr\'evy}
\affiliation{CENBG, UMR 5797 CNRS/IN2P3, B. P. 120, F-33175 Gradignan 
Cedex, France}
\author{D.~Guillemaud-Mueller}
\affiliation{Institut de Physique Nucl\'eaire, IN2P3-CNRS,
F-91406 Orsay Cedex, France}
\author{G.~Hagen}
\affiliation{Physics Division, Oak Ridge National Laboratory,
Oak Ridge, TN 37831, USA}
\affiliation{Department of Physics and Astronomy, University of
Tennessee, Knoxville, TN 37996, USA}
\author{M.~Hjorth-Jensen}
\affiliation{National Superconducting Cyclotron Laboratory and Department of Physics and Astronomy, Michigan
  State University, East Lansing, MI 48824-1321, USA}
\affiliation{Department of Physics and Center of Mathematics for Applications, University of Oslo, N-0316 Oslo, Norway}
\author{F.~Ibrahim}
\affiliation{Institut de Physique Nucl\'eaire, IN2P3-CNRS,
F-91406 Orsay Cedex, France}
\author{A.~Kerek}
\affiliation{Department of Physics, Royal Institute of Technology, 
SE-10691 Stockholm, Sweden}
\author{A.~Krasznahorkay}
\affiliation{Institute for Nuclear Research,
Hungarian Academy of Sciences, P.O. Box 51, Debrecen, H-4001, Hungary}
\author{M.~Lewitowicz}
\affiliation{Grand Acc\'el\'erateur National d'Ions Lourds, CEA/DSM-CNRS/IN2P3, 
B.P. 55027, F-14076 Caen Cedex 5, France}
\author{S.~M.~Lukyanov}
\affiliation{FLNR, JINR, 141980 Dubna, Moscow region, Russia}
\author{ S.~Mandal}
\affiliation{Gesellschaft f\"ur Schwerionenforschung, D-64291 Darmstadt, Germany}
\author{P.~Mayet}
\affiliation{Gesellschaft f\"ur Schwerionenforschung, D-64291 Darmstadt, Germany}
\author{J.\ Mr\'azek}
\affiliation{Nuclear Physics Institute, AS CR, CZ 25068, Rez, Czech Republic}
\author{F.\ Negoita}
\affiliation{IFIN-HH, P. O. Box MG-6, 76900 Bucharest-Magurele, Romania}
\author{Yu.-E.\ Penionzhkevich}
\affiliation{FLNR, JINR, 141980 Dubna, Moscow region, Russia}
\author{Zs.\ Podoly\'ak}
\affiliation{Department of Physics, University of Surrey, Guildford, GU2 7XH,
United Kingdom}
\author{P.\ Roussel-Chomaz}
\affiliation{DSM/IRFU, CEA, F-91191 Gif sur Yvette Cedex, France}
\author{M.G.\ Saint-Laurent}
\affiliation{Grand Acc\'el\'erateur National d'Ions Lourds, CEA/DSM-CNRS/IN2P3, 
B.P. 55027, F-14076 Caen Cedex 5, France}
\author{H.\ Savajols}
\affiliation{Grand Acc\'el\'erateur National d'Ions Lourds, CEA/DSM-CNRS/IN2P3, 
B.P. 55027, F-14076 Caen Cedex 5, France}
\author{G.\ Sletten}
\affiliation{Niels Bohr Institute, University of Copenhagen, 2100 Copenhagen, Denmark}
\author{J.\ Tim\'ar}
\affiliation{Institute for Nuclear Research,
Hungarian Academy of Sciences, P.O. Box 51, Debrecen, H-4001, Hungary}
\author{C.\ Timis}
\affiliation{IFIN-HH, P. O. Box MG-6, 76900 Bucharest-Magurele, Romania}
\author{A.\ Yamamoto}
\affiliation{Department of Physics, University of Surrey, Guildford, GU2 7XH,
United Kingdom}

\begin{abstract}
The structure of the nucleus $^{25}_{\,\,9}$F was investigated
through in-beam $\gamma$-ray spectroscopy of the fragmentation of
$^{26}$Ne and $^{27,28}$Na ion beams. Based on the particle-$\gamma$ and
particle-$\gamma\gamma$ coincidence data, a level scheme was
constructed and compared with shell model and coupled-cluster
calculations.  Some of the observed states were interpreted as
quasi single-particle states built on top of the closed-shell nucleus $^{24}$O, while the others
were described as states arising from coupling of a single proton 
to the 2$^+$ core excitation of $^{24}$O. 
\end{abstract}

\pacs{23.20.Lv, 25.70.Mn, 27.30.+t, 21.60.Cs}
\date{\today}
\maketitle

\section{Introduction}

Nuclei in the vicinity of doubly-closed-shell nuclei are expected to
exhibit simple shell model structures, which can often be interpreted
in terms of selected single-particle degrees of freedom built on a
closed-shell core.  For stable nuclei, several such cases are known in
the vicinity of, for example, nuclei like $^{16}$O, $^{40}$Ca,
$^{56}$Ni, $^{132}$Sn and $^{208}$Pb show such characteristics which is 
relatively well understood. However, more complex structures were observed 
for short-lived isotopes with extreme N/Z ratios.

Recent experiments indicated that $N=16$ was a magic number
close to the neutron drip line. This was first evidenced by neutron 
separation energy and cross section measurements
\cite{ozawa}. Additional information came from the spectroscopy of the
$N=16$ isotones. The energy spectrum of $^{26}$Ne resembled that of a vibrator 
nucleus \cite{Lepailleur}, while only unbound excited states above S$_n$=4.09(10)
MeV\cite{oxygen}, namely a $J^\pi =2^{+}$ state at 4.7 MeV and a $J^\pi= 1^{+}$ at 5.3
MeV~\cite{24ohoffman, 24otshoo} were found in $^{24}$O.  Moreover, a small quadrupole
deformation parameter, $\beta_2=0.15$~\cite{24otshoo}, as well as a large $s$ wave spectroscopic factor of 1.74$pm$0.19 \cite{kanungo} was determined
for the ground state of $^{24}$O, providing further support for its closed N=16 neutron shell. 
It was also proven that the drip line was reached at $^{24}$O for the 
oxygen isotopes reflected in the neutron separation energy step of 4.8 MeV 
between $^{24}$O and $^{25}$O \cite{25o}.  A compatible
value of 4.95 MeV was obtained from the $2J+1$ weighted average energy
of the two $J^\pi=2^{+}$ and $J^\pi=1^{+}$ states in $^{24}$O~\cite{24ohoffman}.

Adding a single valence proton to the doubly-closed-shell nucleus
$^{24}$O, $^{25}$F is expected to have a rather simple structure. Its
energy spectrum can be described up to the neutron separation energy
of 4.36(12)~MeV~\cite{sn} as a few single-proton states coupled to
the ground and first excited states of $^{24}$O. Deviations from this
straightforward picture may arise from the following: 
$^{25}$F is expected to be located at the frontier of emerging new
structures induced by intruder configurations, possibly leading to
cluster configurations at low excitation energy. Recent cluster-model
calculations \cite{kimura} showed that the energy of cluster states
associated with proton cross shell excitations increased with
increasing neutron number among the fluorine
isotopes. Conversely, the energy of a new class of intruder cluster
configurations associated with coherent proton and neutron cross shell
excitations decreased when approaching the neutron drip
line. According to these calculations all the intruder states 
appeared above the neutron separation energy in $^{25}$F \cite{kimura}.  The
lowest-lying proton intruder state is predicted by shell model calculations using the WBP interaction in the $spsdpf$ space at
about 4.3 MeV in the $^{25}$F nucleus \cite{frank,smith}. Recently,
two neutron-unbound states have been reported at 28(4)~\cite{frank} and
300(170)~\cite{smith} above the neutron separation energy which are expected to have intruder $J^\pi= {1/2}^-$ and
intruder $J^\pi={3/2}^-$ or normal $J^\pi=5/2^+$ character, respectively.

The coupling of the weakly bound states to the continuum might also perturb 
the structure of $^{25}$F. All the states
arising from the coupling of a proton to the excited states of
$^{24}$O are expected to lie close to the neutron separation
energy. The first excited state of $^{24}$O itself is unbound by about
1~MeV~\cite{24otshoo}. The effects of continuum coupling were shown 
in the description of the properties of neutron-rich oxygen isotopes in
Refs.\cite{Tsuki,Volya}. Coupling to the
non-resonant continuum and/or to eventual low-lying resonances may
also play an important role in explaining the excitation energy spectrum
of $^{25}$F.


Preliminary papers on the study of $^{25}$F from the
fragmentation of the stable $^{36}$S beam at {\sc Ganil} was already published
\cite{prelim1,prelim2,prelim3}. We reported on four $\gamma$ lines, 
two of which was confirmed in an experiment performed at 
RIKEN~\cite{riken}. In the final analysis of the spectra 7 $\gamma$ rays could be assigned to $^{25}$F \cite{APP}, however, $\gamma$-$\gamma$ coincidence was
not available and only a tentative level scheme could be
constructed for $^{25}$F from these data. In the present paper we
show the results obtained on $^{25}$F at {\sc Ganil} by a
detailed in-beam $\gamma$-spectroscopic study via double-step
fragmentation reaction. The experimental analyzes are
interpreted in terms of both shell-model and
coupled-cluster (CC) calculations. The effective Hilbert space for the
shell-model calculations is defined by the $1s0d$
shell. CC calculations involve much larger effective
Hilbert spaces, typically ten or more major oscillator shells. For
states close to the separation energy one expects that correlations
from states in the continuum may play a larger role, suggesting
thereby the need for larger Hilbert spaces.  The larger
dimensionalities mean however that only selected correlations are
summed to infinite order in CC approaches, in contrast to
configuration interaction calculations performed by the nuclear
shell-model. For the latter, the many-nucleon eigenvalue problem is solved
numerically exactly in a limited space.  The degrees of freedom of say
the $1s0d$ shell studied here, may however not capture the relevant
physics of more neutron-rich fluorine isotopes. These aspects will be
discussed in our theoretical analysis of the experimental data.

\section{Experimental methods}

In the double-step fragmentation reaction a primary beam of $^{36}$S
at 77.5~MeV$\cdot$A with a mean intensity of 400~pnA was delivered by
the two \textsc{Ganil} cyclotrons to induce fragmentation reactions in
a carbon target of 348~mg/cm$^{2}$ thickness placed in the
\textsc{Sissi} \cite{sissi} device. The produced nuclei were selected by means of
the \textsc{Alpha} spectrometer equipped with a 130~mg/cm$^{2}$ Al
wedge at the dispersive focal plane.  The magnetic rigidity of the
spectrometer and the optics of the beam line were optimized for the
transmission of secondary beam particles with $N/Z\approx5/3$ composed
of $^{24}$F, $^{25,26}$Ne, $^{27,28}$Na and $^{29,30}$Mg nuclei with
energies varying from 54~MeV$\cdot$A to 65~MeV$\cdot$A. These nuclei
subsequently impinged on an 'active' target, made of a plastic
scintillator (103~mg/cm$^{2}$) sandwiched by two carbon foils of
51~mg/cm$^{2}$ each, placed at the entrance of the \textsc{Speg}
spectrometer \cite{speg} to induce a secondary reaction.  The plastic scintillator
part of the 'active' target was used to identify the incoming nuclei
through energy loss and time-of-flight measurements. A total of
3$\cdot$10$^4$ $^{25}$F nuclei were produced in the fragmentation
reaction, mainly from $^{26}$Ne and $^{27,28}$Na secondary beams.  The
$^{25}$F nuclei were selected by the \textsc{Speg} spectrometer and
identified at its focal plane by a combined use of energy loss,
total energy, time-of-flight and focal-plane position
information. This latter parameter served to correct the
time-of-flight value from the various flight path lengths of the
fragments in the spectrometer operated in a dispersive
mode. The energy losses and positions of the fragments were determined
by the use of an ionization (70~cm long) chamber and a set of two x-y drift
chambers (12~cm thick and 80~cm width each). The residual energy of the fragments
was measured in a 2~cm thick plastic scintillator whose timing signal
served to determine the time-of-flight (T$_{pl}$)
with respect to the cyclotron radio frequency and to the plastic
scintillator of the 'active' target.

An array of 74 BaF$_2$ crystals was mounted in two hemispheres around
the 'active' target at a mean distance of 21~cm, covering a total
solid angle of 80\%. It was used to detect in-flight $\gamma$-rays
emitted with $v/c \simeq 0.34$ in coincidence with the $^{25}$F
fragments. The $\gamma$ spectra were corrected for the Doppler-shift,
yielding a full width at half maximum of about 12~\%.  The BaF$_2$
array efficiency was calibrated between 120~keV and 1410~keV using
$^{137}$Cs and $^{60}$Co standard $\gamma$ sources placed at the
target position. A typical photo-peak efficiency of about 20~\% is
obtained at 1.3~MeV.  Owing to the compact geometry of the array, high
energy $\gamma$ rays could scatter from one detector to another. To
reduce the background and to achieve a higher efficiency at high
energies, an add-back algorithm was used for the events where at
least 2 neighboring detectors fired at the same time.

\section{Experimental results}

\begin{figure}
\begin{center}
\epsfig{width=8.5cm, file=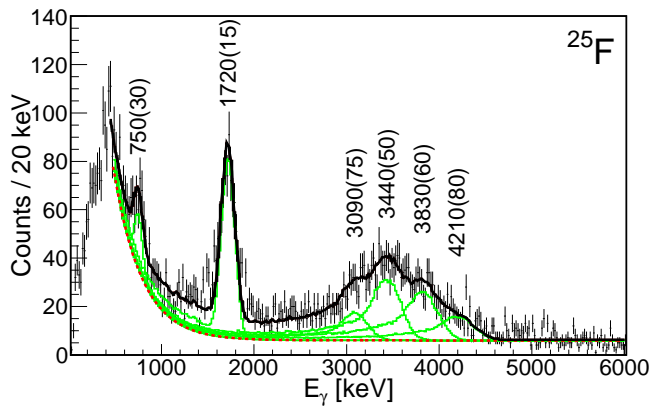}
\caption{Decomposition of the $\gamma$-ray spectrum of $^{25}$F. The
solid black line shows the final fit which includes the pesponse functions
from \textsc{Geant4} simulation(green solid curves) and the additional
exponential background plotted as dashed red line.}
\label{25fsp2}
\end{center}
\end{figure}

The Doppler-corrected $\gamma$-ray spectrum obtained for the $^{25}$F
nucleus is shown in Fig.~\ref{25fsp2}. Two peaks can be seen at 750(30)~keV
and 1720(15)~keV, while a broader structure including several
$\gamma$ lines is present between 3~MeV and 4.5~MeV. This observation is
in agreement with the results obtained in the fragmentation of
$^{36}$S beam \cite{prelim1,prelim2,prelim3,APP}.

The decomposition of the broad structure into individual peaks
requires the determination of the energy-dependent $\gamma$ width of
the peaks, which was obtained from single $\gamma$ peaks observed in
other reaction channels in the same experiment. An almost linear
energy dependence of the peak width was observed.
The response function of the BaF$_{2}$ crystals leads to a $\gamma$
peak with a low-energy tail, the energy-dependent shape of which was
simulated by use of the \textsc{Geant4} package. This low-energy tail
is due to single and double escape as well as the Compton events which
remain after the Compton suppression treatment.  The line shape
obtained with the simulation was tested successfully in the case of
$^{22}$O \cite{oxygen}.  The fitting of the broad structure in the
$^{25}$F spectrum was made using the deduced $\gamma$ line shape and
an exponential background, yielding $\gamma$-rays with energies of
3090(75), 3440(50), 3830(60) and 4210(80)~keV as shown in
Fig.~\ref{25fsp2}. These energies are in agreement with the values
deduced after reevaluation of the spectra obtained in the single step
fragmentation~\cite{APP}.

\begin{figure}
\begin{center}
\epsfig{width=6cm, file=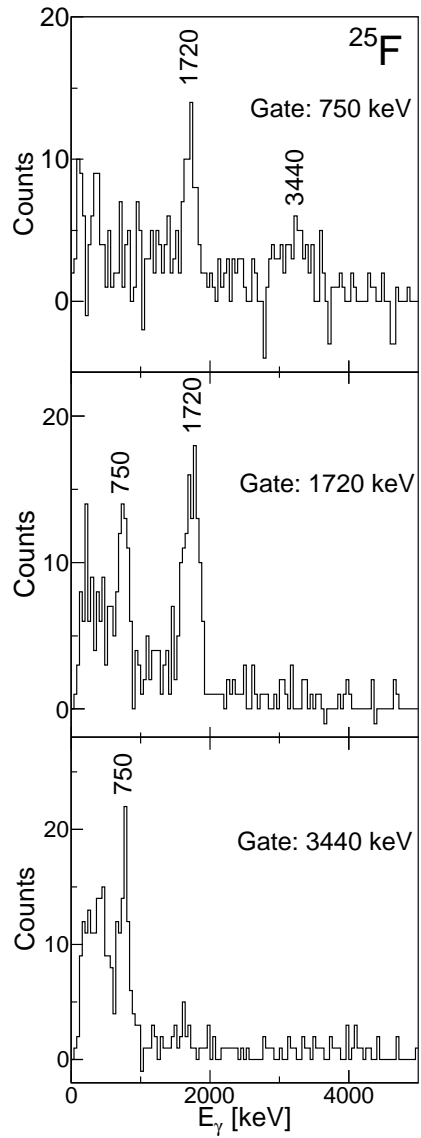}
\caption{Spectra of $^{25}$F from $\gamma\gamma$ coincidence  using the 750,
  1720 and 3440~keV transitions as gates.}
\label{25fko}
\end{center}
\end{figure}

For the level scheme construction, $\gamma\gamma$-coincidence
matrices were created.  The most intense 1720 keV
$\gamma$ line was found to be in coincidence with itself, as well
as with the 750 keV line. Furthermore, the 750 keV transition is in
mutual coincidence with a 3440 keV $\gamma$ ray as it can be seen in
Fig.~\ref{25fko}. In the analysis of the coincidence spectra the same line 
shapes and exponential bacground was assumed as for the single spectra.

The level scheme of $^{25}$F was established by using
$\gamma\gamma$-coincidence matrices as well as energy and intensity
balances. The energy and uncertainty of the excited states given below
were determined taking into account the energies and uncertainties of all
the $\gamma$-rays connected to a given state via the fitting procedure of
the \textsc{Radware} package~\cite{radware}. 

The coincidence relation of the 1720~keV $\gamma$
line with itself suggests two $\simeq$1720~keV $\gamma$ rays in
cascade, establishing stats at about 1720~keV and
3400~keV. As this energy overlaps with that of the 3440(50) keV transition 
within the experimental uncertainties, a level is proposed at 3440(21) keV.
The $\gamma$ line of 750~keV was chosen to feed the 3440~keV level since it was 
in coincidence with both $\gamma$ rays of 1720~keV and 3440~keV. 
The sum 1720(15)+1720(15)+750(30)$\simeq$4190(60)~keV overlaps with the
energy of the 4210(80)~keV $\gamma$-ray establishing a state at 4195(35)~keV. 
The remaining transitions at 3090(75) and 3830(60)~keV, which were not observed 
in coincidence with any transitions, were placed to directly feed the ground state,
establishing levels at the corresponding energies. A weak 2140(30) keV line 
was observed in the single step fragmentation reaction \cite{APP}, the intensity 
of which was enhanced in the multiplicity 2 spectrum, suggesting that it was in
coincidence with another transition. Energetically such a
transition could connect the 1720~keV and 3830~keV states. In the
single step reaction the 3830 keV line was the strongest, which explains
why this transition was below the observation level in the present work.

\begin{figure}
\begin{center}
\epsfig{width=8.5cm, file=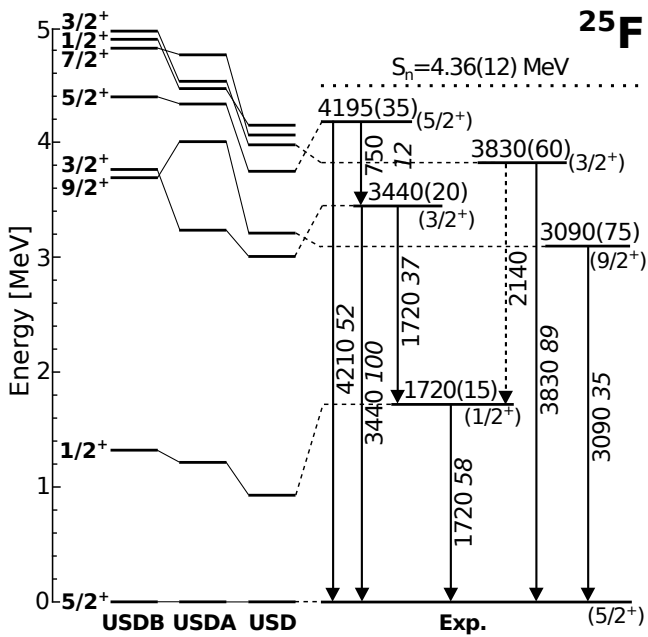}
\caption{Proposed level scheme of $^{25}$F compared to the shell model
  calculations performed using the \textsc{usd}, \textsc{usda} and
  \textsc{usdb} interactions.  Energies are given along the transitions as well as their 
  relative intensities in italics. The uncertainties of the relative intesities 
  are below 20\%. The 2140 keV line has been adopted from Ref.~\cite{APP}.}
\label{25fls}
\end{center}
\end{figure}

The relative intensities of the 750(30)~keV, 3090(75)~keV, 3440(50)~keV, 3830(60)~keV
and 4210(80)~keV transitions were determined on the basis of
energy-dependent $\gamma$-ray efficiencies. The situation was more
complex for the unresolved 1720~keV doublet, since the intensity of
its members could not be derived directly.  
First the intensity ratio of the higher-lying member of the 1720~keV
doublet and of the 3440 keV $\gamma$-ray deexciting from the
3440(20)~keV state was derived in a spectrum gated with the $\gamma$ ray of 750~keV. 
From this ratio the
intensity of the higher-lying member of the 1720~keV doublet was
deduced using the corresponding peak areas in the single spectrum. The
intensity of the lower-lying member of the doublet was obtained by
subtracting the intensity of the higher lying component from the
integral of the whole 1720~keV intensity. The resulting
intensities of the $\gamma$ transitions are presented in
Fig.~\ref{25fls}.  

\section{Discussion}

The established level scheme of $^{25}$F (Fig.~\ref{25fls}) is
compared to the results of shell-model calculations using effective
interactions like the \textsc{usd}~\cite{usd}, \textsc{usda} and
\textsc{usdb} \cite{usda} interactions defined for the $1s0d$ valence
space. From a naive single-particle picture we expect that the ground state of
$^{25}$F has spin and parity $J^{\pi}=5/2^+$. It corresponds to the
filling of the $d_{5/2}$ single proton orbit above the $^{24}$O core.

The first excited state at 1720(15) keV is supposed to
correspond to the $J^{\pi}=1/2^+$ state. In the simple single-particle
picture, the $1/2^+$ state can be interpreted as a proton excitation
to the $1s_{1/2}$ orbit.  All other excited states, predicted above
3~MeV in the $sd$ space, originate from the coupling of a proton 
single-particle state
to excitations of the $^{24}$O core in addition to the proton $d_{3/2}$ 
excitation.  In addition, negative parity intruder or cluster states arising 
from a wider model space may also be present.

Considering that the state at
3440~keV decays both to the ground state of $J^{\pi}=5/2^+$ and the
first excited $J^{\pi}=1/2^+$ state, a tentative $J^{\pi}=3/2^+$
assignment can be given to this state. Similarly, the 4195 keV state
decays to the $J^{\pi}=5/2^+$ ground state and to the $J^{\pi}=3/2^+$
excited state, but not to the $J^\pi=1/2^+$ state, suggesting a
$J^{\pi}=5/2^+$ assignment for the initial state. The $\gamma$
branching ratios are also in a qualitative agreement with the
predictions of the $sd$ shell model. While comparable intensities are expected for the
transitions from the $J^\pi=3/2^+_1$ state to the $J^\pi=5/2^+_1$ and $J^\pi=1/2^+_1$
ones, the M1 transition from the $J^{\pi}=5/2^+_2$ to the
$J^{\pi}=5/2^+_1$ state is strongly hindered. This latter transition
could only be observed since the $E_{\gamma}^3$ energy factor gives an
enhancement of a factor of about 200. 

All the shell-model interactions
predict a $J^{\pi}=9/2^+_1$ state below the $J^{\pi}=5/2^+_2$ level.
A good candidate for this $J^{\pi}=9/2^+_1$ level is the level
derived at 3090(75)~keV.  On the basis of shell model calculations and
the fact that it directly decays to the ground state, the 3830~keV its 
spin-parity can be limited to $1/2^+ - 9/2^+$. Considering that the 2140(30)~keV 
line observed in the single step reaction can also be assigned to the 
decay of the 3830~keV state, a spin 3/2 might be assigned to this state. 
The branching ratios deduced from the single step reaction are consistent 
with the branching ratios calculated in the shell model for the $3/2_2$ state. 

In Fig.~\ref{25fls}, we can see that rather strong deviations can be found
between the different versions of the shell-model interactions.  The
\textsc{usd} interaction fits reasonably well the experimental data,
while the \textsc{usda} and \textsc{usdb} interactions stretch the
energy spectrum too much. Assuming the above tentative spin assignments, 
the higher energy group of the experimental states lies about 500 keV below 
the predicted ones. This energy difference is similar to what was observed in 
$^{24}$O. 
Since most of these states originate from the coupling 
of the proton single particle states to the 2$^+$ and 1$^+$ states of $^{24}$O 
the deviation 
might come from the inaccurate 
description of the $^{24}$O core excitation. In this context we mention 
that these states of $^{24}$O have a $\nu s_{1/2}d_{3/2}$ dominant 
configuration, 
where the $1s_{1/2}$ neutron is excited to the unbound $0d_{3/2}$ orbit. In 
this connection it may be interesting to mention that to describe the new data 
-- among others -- the neutron d3/2 single particle energy was increased by 330 
and 440 keV in the \textsc{usda} and \textsc{usdb} interactions, respectively, 
relative to the \textsc{usd} parametrization.

The unbound $d_{3/2}$ neutron may have an extended spatial distribution, 
which can result in a decrease of the interaction matrix elements in which 
the $d_{3/2}$ neutron is involved. Decreasing the $\nu s_{1/2}d_{3/2}$ 
interaction strength results in a decreased splitting of the $\nu s_{1/2}d_{3/2}$ 
doublet, yielding an increase in energy of the 2$^+$ state in $^{24}$O. Decrease 
of the $\pi d_{5/2}\nu d_{3/2}$ interaction causes the decrease of the 
splitting 
of the $\pi d_{5/2}\oplus 2^+$ multiplet. As a result, both effects would 
worsen the agreement with experiment, thus the spatial extension of the unbound 
neutron $0d_{3/2}$ orbit cannot explain the observed deviation. A possible explanation 
may be the coupling to continuum configurations, as proposed in \cite{Tsuki}.

A low-energy bound intruder state with spin $J^{\pi}=1/2^-$ was
predicted around the neutron separation energy at 4296~keV using the
$0p1s0d$ model space \cite{frank}.  If bound, such a state would decay
via an $E1$ transition, predominantly to the $J^{\pi}=1/2^+$ state.
Experimentally, we do not observe any state with such decay property
and energy, suggesting that the first intruder state in $^{25}$F is
unbound in agreement with the results of Franck {\it et al.}~\cite{frank}.

We also performed microscopic coupled-cluster~\cite{cc} (CC) calculations~(see~\cite{shavittbartlett} 
for a textbook presentation) and compared to experimental results. The coupled-cluster 
method is ideally suited to calculate
properties of closed-shell nuclei and their immediate neighbors. While the
shell-model restricts the model space to all possible determinants constructed
from a small set of single-particle orbitals around a closed shell core, the CC
method restricts the number of particle-hole excitations allowed in the
determinant basis but uses a large single-particle space. Such particle-hole
correlations are summed to infinite order depending on the level of
approximation. Coupled-cluster theory is thus a non-perturbative method.

To describe ${}^{25}$F, that has a proton attached to ${}^{24}$O, we use
particle-attached equation-of-motion coupled-cluster
theory~(PA-EOM-CC)~\cite{gour2006,pa}. In PA-EOM-CC the ${}^{25}$F wave function
is described by a linear expansion of $1$p, $2$p-$1$h, $3$p-$2$h\ldots
excitations on top of the ground state of ${}^{24}$O. In our earlier
applications, we truncated the expansion at the $2$p-$1$h level, which works
particularly well for low-lying states that are dominated by $1$p excitations
from a closed-shell ground state~\cite{gour2006,pa,f17,o25}.

In this work we are also interested in describing excited states in ${}^{25}$F
that can be viewed as a proton attached to the $J^\pi = 2^+$ excited state in
${}^{24}$O. Clearly, to describe such states we need to go beyond the $2$p-$1$h
truncation level.  As a rule of thumb, the level of approximation should be one
order more than what is considered important for the state under study. 
We therefore include $3$p-$2$h configurations as presented in Ref.~\cite{pa3p2h}, but as a
first approach we only include the terms that determines the
$3$p-$2$h amplitudes in the coupled-cluster singles and doubles~(CCSD)
approximation.
The similarity transformation generated by solving the CC equations
induce additional many-body terms in the Hamiltonian. Only diagrams that contain
at most two-body parts of the Hamiltonian have been included at the $3$p-$2h$
level, while all diagrams have been included up to the $2$p-$1$h level.

We use interactions from chiral effective field theory at two
different orders.  First we look at the newly optimized chiral
interaction at next-to-next-to-leading~(N${}^2$LO) order from
Ref.~\cite{Ekstrom2013}. Already at this order in the chiral
expansion, three-body diagrams appear, but these have not been
included here. This interaction resulted in an excellent agreement for
both binding energies and selected excited states for oxygen
isotopes. Second, we look at the chiral interaction at
next-to-next-to-next-to-leading~(N${}^3$LO) order from
Ref.~\cite{Entem2003}. Here, the effects of three-nucleon forces are
treated as density-dependent corrections to the nucleon-nucleon
interaction by integrating one nucleon in the leading-order chiral
three-nucleon force over the Fermi sphere with a Fermi momentum $k_F$
in symmetric nuclear matter~\cite{eff3nf}. We use the parameters
already established in Ref.~\cite{Hagen2012a}. These parameters were
used recently in a study of oxygen isotopes \cite{Hagen2012a}, and
also for ${}^{26}$F in Ref.~\cite{Lepailleur2013}.

Due to the increased computational cost of including $3$p-$2$h configurations,
the single-particle space is limited to a Hartree-Fock basis, built
from $N=10$ major harmonic oscillator shells. This is not large enough for
the total binding energies to be converged, but the relative energies
exhibit a much faster convergence as function of the number of
shells. The N${}^2$LO interaction is rather soft, meaning that the
relative energies are practically converged at this level. On the other hand our
results using the N${}^3$LO interaction exhibits a slower convergence rate,
meaning that a larger model space is needed.

\begin{figure}
    \centering
    \epsfig{width=0.45\textwidth, file=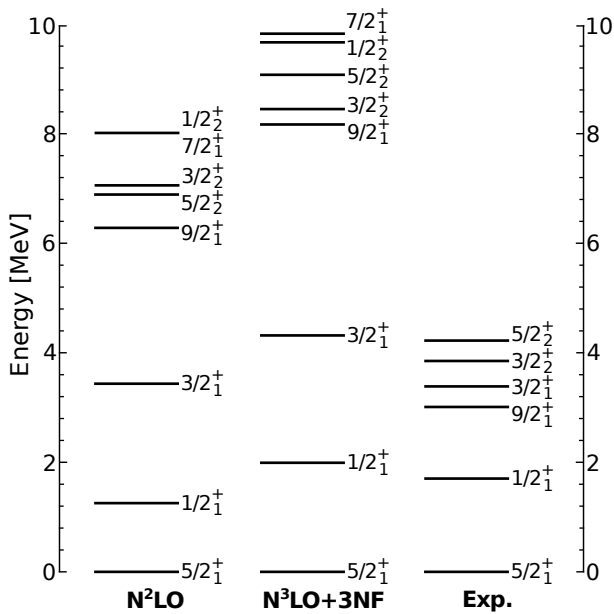}
    \caption{\label{fig:levels} Relative energies of ${}^{25}$F calculated using
    coupled-cluster theory with two different interactions compared with
    experiment. The label N${}^2$LO refers to the optimized chiral interaction at third
    order\cite{Ekstrom2013}, while the interaction labeled N${}^3$LO+3NF is the
    chiral interaction of Ref.~\cite{Entem2003} where the effect of three nucleon
    forces are treated as a density-dependent nucleon-nucleon force~(See text
    for details).}
\end{figure}

In Fig.~\ref{fig:levels} we plot the relative energies of selected positive
parity states in ${}^{25}$F calculated with the two different interactions, as well as
experimental energies. The first excited $J^\pi = {1/2}^+$ and $J^\pi = {3/2}^+$ states
are reproduced reasonably well by both interactions, but the remaining states are at too
high excitation energies as
compared to data. This deviation might suggest that for these states, the 
current level of
approximation is not good enough. Including $4$p-$3$h configurations, where the latter 
account for $3$p-$3$h excitations of the ${}^{24}$O reference state, may  further
reduce the relative energies by $2-4$~MeV. This guess is based on our
observation that the relative energies for the high-lying states changes by
10-20 MeV by going from the $2$p-$1$h to the $3$p-$2$h truncation level.
To get a better insight into which contributions are important, we examine here in more detail
the structure of the calculated wave functions.

\begin{table}
    \caption{\label{tab:norms}Partial norms of wave functions using up to 3p2h amplitudes on top of the ${}^{24}$O reference state. The interaction is the newly optimized chiral interaction at third order~(N${}^2$LO) \cite{Ekstrom2013}.}
    \begin{ruledtabular}
\begin{tabular}{cccc}
     & 1p & 2p1h & 3p2h \\
    \hline
    ${5/2}^+_1$ & 0.63 & 0.30 & 0.07 \\
    ${1/2}^+_1$ & 0.56 & 0.36 & 0.08 \\
    ${9/2}^+_1$ & 0.00 & 0.74 & 0.26\\
    ${3/2}^+_1$ & 0.47 & 0.42 & 0.11\\
    ${3/2}^+_2$ & 0.01 &0.72 & 0.27 \\
    ${5/2}^+_2$ & 0.01 & 0.73 & 0.26\\
    ${1/2}^+_2$ & 0.03  & 0.72 & 0.25 \\
    ${7/2}^+_1$ & 0.00 & 0.73 & 0.27\\
\end{tabular}
\end{ruledtabular}
\end{table}

Table~\ref{tab:norms} lists the partial norms of the
PA-EOM-CC wave functions calculated with the optimized N${}^2$LO
interaction. They sum up the parts of the wave function in $1$p,
$2$p-$1$h and $3$p-$2$h configurations where the sum of all partial
norms is one. We list only the partial norms for this N$^2$LO
interaction since those obtained with N$^3$LO interaction with density
dependent three-body force produce qualitatively similar results.

We
see that the three presumed single-particle states have a large~(30-40\%) 
contribution from $1$p-$1$h excitations of the ${}^{24}$O
reference state. If these were predominantly single-particle states,
we would expect a contribution of approximately ten percent or less
from $2$p-$1$h configurations. As it turns out, $2$p-$1$h
configurations are not enough to describe these states and they are
not converged before $3$p-$2$h configurations are included. We see
that roughly ten percent of the wave functions for these three states
come from $3$p-$2$h configurations, which implies a good level of convergence at
this order.  This is consistent with the
shell-model calculations performed for these states. For the
$J^\pi={5/2}^+$ ground state, the different \textsc{USD} interactions discussed
above give wave functions with a 70-80~\% overlap with single-particle
configurations, whereas the corresponding coupled-cluster numbers are
60-70~\% .
For the first excited $J^\pi={1/2}^+$ and $J^\pi={3/2}^+$
states, the single-particle content of the wave functions has dropped
significantly in both models.

For the three lowest lying positive parity states
$J^\pi={5/2}_1^+$, $J^\pi={3/2}_1^+$ and $J^\pi={1/2}_1^+$ we note
that the admixture from $3$p-$2$h configurations do not exceed some
$10\%$, meaning that we can interpret these states mainly in terms of
$1$p and $2$p-$1$h configurations. The relative energies of these
states are also well-converged within the chosen set of
configurations.

We notice also that for other positive parity states
like the $J^\pi={9/2}_1^+$, $J^\pi={3/2}_2^+$ and $J^\pi={5/2}_2^+$
states, there are considerable admixtures from $3$p-$2$h and more
complicated configurations. Since additional $4$p-$3$h configurations
are expected to reduce the relative energies by $2-4$ MeV, these
changes are most likely larger than contributions that can be obtained
from continuum effects. In Refs.~\cite{Tsuki,Hagen2012a} we found that the
coupling to the continuum in $^{24}$O gave a reduction in relative
energy of 300-500 keV for states close to the neutron separation
energy, just the energy difference found between the experimental 
results and the \textsc{usda/b} calculations.  As suggested by Table~\ref{tab:norms}, a suitable 
description of the high energy spectrum of $^{25}$F requires to treat contributions 
above the (already large dimensionality) $3$p-$2$h configurations. As this treatment 
is beyond the present computational limitations, a discussion of the continuum effects on these states
has to be deferred to a later work.
 
We have also studied negative parity states within the PA-EOM-CC framework. We
find a $J^\pi={3/2}^-$ state (Not shown in Fig.\ref{fig:levels}) at 
approximatly $10$~MeV. It is identified as a spurious $J^\pi = 1^-$ 
center-of-mass excitation~\cite{com3} built on
the $J^\pi={5/2}^+$ ground-state configuration and is therefore not considered a
physical state~\cite{pa,com2}. At around $12$~MeV we find a $J^\pi= {1/2}^-$ 
state that is consistent with a physical state, but a large $3$p-$2$h component 
suggests that it is not yet converged at this level of approximation.

\section{Conclusion}

We studied the structure of $^{25}$F by the use of the in-beam $\gamma$
spectroscopy technique from a double-step fragmentation reaction. Utilizing the
$\gamma$-spectroscopic information, we constructed a level scheme for
$^{25}$F, including the states corresponding to the coupling to the core excitation
of $^{24}$O. Shell-model calculations using parametrized interactions for the
$1s0d$ shell account for many of the  general features of the energy spectrum, 
although a 500 keV deviation from the experiment was found in the
\textsc{usda/b} calculations for the temporarily assigned states of the $\pi
d_{5/2}\oplus 2^+$ multiplet. All the states observed could be interpreted 
within the $sd$ shell model, no bound negative parity intruder or cluster state 
has been found.

We  also performed coupled-cluster calculations
using recent interaction models from effective field theory. These calculations
resulted in a very good agreement with experiment for the low-lying $J^\pi=5/2^+$,
$1/2^+$ and $3/2^+$ states. The
analysis of the wave functions showed also a qualitatively similar picture in
terms of single-particle states as the shell-model calculations.
For higher-lying states that could be assigned to the $\pi
d_{5/2}\oplus 2^+$ multiplet, our coupled-cluster calculations yielded excited
states at too high excitation energy.
An analysis of the structure of the
wave functions indicated that correlations beyond the current truncation level
might be necessary.
In $^{24}$O  we found that the coupling to the continuum gave a
reduction in relative energy of 300-500 keV for states close to the neutron
separation energy, just the energy difference found between the experimental
results and the \textsc{usda/b} calculations.  However, due to the large
dimensionalities introduced by $3$p-$2$h configurations in this work, we were
not able to provide a proper estimate of continuum effects in $^{25}$F.
  
\acknowledgments
{\small This work was partly supported by the European Union's Seventh Framework
Program under grant agreement no 262010, by a grant of the Romanian 
National Authority for Scientific Research, CNCS$-$UEFISCDI, project number
PN-II-ID-PCE-2011-3-0487, and also by OTKA contract number K100835 and 
NN104543, PICS(IN2P3) 1171, INTAS 00-00463, GACR 202-04791, RFBR N96-02-17381a 
grants, the Bolyai J\'anos Foundation and the 
T\'AMOP-4.2.2/B-10/1-2010-0024 project. The T\'AMOP project is
co-financed by the European Union and the European Social Fund.
In addition, this work was partly supported by the Office of Nuclear
Physics, U.S. Department of Energy (Oak Ridge National
Laboratory), under No.DE-SC0008499 (NUCLEI SciDAC-3 Collaboration), and
the Field Work Proposal ERKBP57 at Oak Ridge National Laboratory.
An award of computer time was provided by the Innovative and Novel
Computational Impact on Theory and Experiment (INCITE) program.  This
research used resources of the Oak Ridge Leadership Computing Facility
located in the Oak Ridge National Laboratory, which is supported by the
Office of Science of the Department of Energy under Contract
DE-AC05-00OR22725 and used computational resources of the
National Center for Computational Sciences, the National Institute for
Computational Sciences and the Notur project in Norway. Support from the Research Council of Norway 
under contract ISP-Fysikk/216699 is acknowledged. BAB acknowledges support from NSF grant PHY-1068217. 
This research was partly realized in the frames of TÁMOP 4.2.4. A/2-11-1-2012-0001 
“National Excellence Program – Elaborating and operating an inland student and researcher personal support system”.
The project was subsidized by the European Union and co- financed by the European Social Fund.


\end{document}